\newcommand{\be}{\begin{equation}}
\newcommand{\ee}{\end{equation}}
\newcommand{\beq}{\begin{eqnarray}}
\newcommand{\eeq}{\end{eqnarray}}

\def\p{\partial }

\def\ov{\over  }

\documentclass[12pt]{article}
\usepackage{epsfig}
\usepackage{graphicx}
\usepackage{epsfig,cite}

\input epsf
\usepackage{amsmath,amsfonts,epsfig,color,latexsym}
\usepackage{amssymb}
\usepackage{amsfonts}
\usepackage{amssymb}
\begin{document}
\vskip1cm

{\centerline{Gravitational collapse of a macroscopic string by a Newtonian description}}
{\centerline{including the effect of gravitational radiation}}

\vskip1cm

{\centerline{Roberto Iengo \footnote{iengo@sissa.it} }}   
{\centerline{SISSA, Trieste and INFN, sezione di Trieste}}

\vskip1cm

{\bf{Abstract.}}  We make an attempt to dynamically study, in four space-time dimensions, 
the classical gravitational collapse of a macroscopic circular fundamental string,
by a truncation of the Einstein equations that suppresses  retarded features but
keeps the main self-gravity peculiarities of the relativistic string dynamics, 
and allows the investigation of a possible infinite red-shift. 
The numerical solution of the string evolution in the self-induced metric 
shows an infinite red shift at a macroscopic size of the string,
when the string reaches the velocity of light.
We further include the back-reaction of the radiation of gravitons 
which induces energy dissipation: now the
velocity of light is not reached, the infinite red shift 
does not form and the string simply shrinks with damped oscillations.

\vskip2cm

{\bf{Introduction and Summary.}} 

\vskip0.5cm

We propose to study, in four space-time dimensions, the gravitational collapse of a 
classical macroscopic fundamental string, with no angular momentum,
which in flat space-time is described by a circular pulsating solution, periodically shrinking to zero size.
One can expect that self-gravity will accelerate the shrinking of the string, leading to a complete
gravitational collapse.

The string dynamics is determined by three kind of forces
\footnote{Here we neglect the coupling of the string to $B_{\mu\nu}$ and to the dilaton.}: 
1) the string tension, which would make the string  
oscillate, 2) the gravitational self-attraction, which would drive the string to zero size and increase its velocity, 
3) the back-reaction due to  the radiation of gravitons, which is a dissipative, friction-like, force.
Forces 2) and 3) include highly non-linear effects.

Self-gravitating strings, including the effect of radiation, have been studied long ago in ref. \cite{Qua},  the 
results having also being reported in the "Cosmic String" textbook by Vilenkin and  Shellard \cite{Vile}.
This work uses linear gravity to compute the metric and to evolve the string iteratively.
In the case considered there, in flat space-time the string size  does not go to zero; it is found that
including the combined effect of gravity and radiative dissipation does not cause a collapse,
rather, the string looses  energy by radiation.

In our case instead, the string initial state is driven toward  shrinking already by the force 1). 
This setting is more similar to the case discussed by Hawing in \cite{Hawk2}: he observed
that, with the linearized theory, the energy carried away by radiation in a collapse up to a point 
seems to be infinite.  Then,  taking as the initial stage a string on a null surface, that is having reached the 
velocity of light, he used geometrical methods to put a lower bound on the horizon and hence
on the final mass, finding an upper bound of 29\% for the energy difference between the initial and the final stage.

We would like to get a dynamical picture of this process. In this paper, as an attempt to begin with, 
we introduce some simplifications in the dynamics, however trying to keep the main physical ingredients
including non-linear features.
We focus on the dynamical rather than on the geometrical aspects of the problem.
In our attempt, we make a truncation by taking $g_{\mu\nu}=(g_{00},-1,-1,-1)$ and  
keeping only the $R_{00}$ component of the Einstein equations.

This gives a kind of  Newtonian description, where retarded effects are not taken into account,
but it keeps into account the relativistic string dynamics and the gravitational role of the string kinetic energy,
 with the peculiarities of the world-sheet constraints and related non-linear effects.
 
Since $g_{00}$ determines the gravitational red-shift (see for instance \cite{Dinv}), 
our truncation provides a modern "Laplace" 
\footnote{Laplace found the radius of a star to be Schwarzschild when the escape velocity equals $c$.}
 criterion for the collapse. It turns out that
automatically $1\geq g_{00}\geq 0$, reasonably implying that we cannot describe the string beyond the 
infinite red-shift  defined by  $g_{00}=0$. In our initial conditions  $g_{00}=1$. 

Let us stress that our intention is not the description of a Black Hole (or a Black Ring or whatever Black),
a word which we carefully avoid, except when we are forced to do referring to the literature.
Our approach can only indicate the possibility of an infinite red-shift, and it does not provide 
theoretical instruments to investigate what happens after it.  

We solve consistently the string evolution in the self-gravitation background metric, 
 regulated with a minimal length $\sqrt{\alpha'}$, to implement in a simple way 
 the absence of UV-divergences in String Theory. 

The string dynamics appears to be rather peculiar. The evolution equation is almost scale invariant, 
essentially due to the scale invariance of $M/R$ for $M\sim R$; that is, the only dependence on the 
overall size, which is the same as the mass, comes combined with $\alpha'$,
and it is weak, roughly logarithmical.  
Further, the gravitational attraction is controlled by the
kinetic energy (per unit mass), rather than by the mass,
 therefore the string velocity plays a crucial role. 

We find that the numerical program stops at a point where the string is still macroscopic but its  
kinetic energy diverges and the velocity goes to the velocity of light, in agreement with \cite{Hawk2}:
at this point $g_{00}=0$ on the string. That is, the infinite red-shift makes the string disappear
and we can no longer follow its evolution. 

We further include the radiation of gravitons and its back-reaction on the string evolution.
We take the radiation  emission, at each moment of the evolution, to be given
by a standard textbook flat-space-time formula (see \cite{Landau}), 
assuming it to hold in the locally inertial frame of the free-falling the string. 
The main effect of the radiation is to provide self-dampening, 
since the energy dissipation  is strong for strong accelerations.

It turns out that this friction force due to radiation does not depend on the overall size or mass,
that is, it is scale invariant.

The numerical solution shows that the back-reaction of the radiation is important 
before the stage of the string reaching the velocity of light,
which is in fact never reached.
The numerical program does not stop and $g_{00}> 0$ always:
 in a first interval the kinetic energy is large but not divergent, 
 $g_{00}$ on the string becomes small but not zero;
then the string bounces back and it further re-contracts with less large kinetic energy and less
smaller $g_{00}$, and so on. Ultimately, after the initial rather violent phase, the string quietly loses energy as in ref. \cite{Qua}, and shrinks 
indefinitely without catastrophic consequences.
\vskip0.1cm
There are many other cases of extended macroscopic strings in which much of the string periodically concentrates to a very small size, just by the tension force 1) in flat space-time.
In particular, this happens rather generically as the result of the interconnection of two 
rotating strings, see \cite{Russo}.  We expect those cases to be quite similar 
to the case that we have considered.

\vskip0.3cm

There is of course a large literature on themes that are more or less related to our problem. 
 We will quote and comment very few of them, which we find to be more near to some aspects
 of our problem.  Any attempt to completeness would be misleading, and we apologize to the 
 many authors which we miss. 

The gravitational collapse continues to be extensively studied by the astrophysics community, also with
 attention to radiation.  As relevant examples, we quote the pioneer paper
 by Stark and Piran \cite{Piran} dealing with numerical work on the collapse of stars with angular momentum and the more recent  papers on similar cases \cite{Rezzo}, and \cite{Dimme} where, in particular,  
 the results of general relativity are compared with results of Newtonian physics \cite{Zwer}, showing an overall agreement although with some quantitative differences (see also \cite{Dimme2}).
 In those studies of collapse, the gravitational radiation appears to carry a tiny fraction of the total mass.

There is another bunch of papers dealing with the possible formation of Black Holes in scattering  of very energetic 
 initial states, either Black Holes themselves \cite{D'Eath} (and previous work), or elementary particles or wavepackets, see for instance \cite{Giddings},  \cite{Giddings2} and many others dealing with variations on the theme. 
In particular, ref. \cite{D'Eath}  has considered also the gravitational emission during the collision,
 finding a moderate 16\% of the total energy, qualitatively similar to the quoted results of the collapse in astrophysics. 

 One can then ask how it is that in our case gravitational radiation seems to be so important.
 Apart from the obvious possibility that our truncation is not adequate, we may observe again
 the peculiarities of the string with respect to other gravitating bodies: 
 its gravitational potential is controlled by its kinetic energy rather than its mass, and
we find an amplification,
 through non linear terms, of the gravitational effect of the string kinetic energy.
This is due in good part to the role of the world-sheet conformal invariance constraint on the string dynamics.
 Therefore, the collapse strongly accelerates towards high velocities of the string. But at the same time, 
 this acceleration together with the high velocity triggers a strong dissipative radiation, implying
 a strong friction which prevents the string to reach the velocity of light, therefore avoiding from
 being trapped in a horizon.

Other authors have considered self-gravitating fundamental strings, see ref. \cite{Horo} and \cite{Vene}, however from a completely different point of view;
in this case the problem is to study the gravitational effects as a consequence of increasing the string coupling, taking typical string configurations,
more near to a random walk, rather than  the problem of the collapse at fixed coupling of a smooth extended configuration like our initial state.

Finally, last but not least, we mention the work by Amati, Ciafaloni and Veneziano \cite{Ama}, 
an attempt to see whether the scattering of gravitons, as described by  quantum string theory, 
can produce a Black Hole (see also \cite{Ama2}). In this case radiation should be automatically
taken into account, but, as far as we can understand, there is no concluding evidence whether a Black Hole is formed or not.
\vskip3cm
{\bf{Index:}}
\vskip0.5cm
1) Statement of the problem.  \hfill  {pag.5}

2) The truncation.  \hfill { pag.6}

3) Solving the equations. \hfill  { pag.7}

4) Results describing the collapse. \hfill  {pag.7} 

\hfill  (figures pag.8)

5) Including the radiation of gravitons. \hfill {pag.9}

6) Including the radiation: results.  \hfill  { pag.10}

\hfill  (figures pags. 11,12)

7) Conclusions. \hfill {pag.12}

8) References. \hfill {pag.13}

\newpage

{\bf{1) Statement of the problem.}}
\vskip0.1cm
We would like to study the gravitational collapse of a macroscopic closed circular string, 
which can be parametrized as
\be
X(\sigma,\tau)=R_0 f(\tau)Cos(\sigma) ~~~ Y(\sigma,\tau)=R_0 f(\tau)Sin(\sigma) ~~~ 
X^0(\sigma,\tau)=R_0 g(\tau)
\label{X}
\ee
In flat spacetime we get the equation
$$
f''+f=0
$$
and we take the solution $f(\tau)=Cos(\tau)$ representing a circular pulsating  string,
which has a maximal radius for $\tau=n\pi$ and shrinks to zero for $\tau=\pi/2+n\pi$.
The Virasoro costraint fixes $g'(\tau)=\sqrt{(f')^2+f^2}$ : indeed 
$(\p_\tau X^0\pm\p_\sigma X^0)^2=(\p_\tau\vec X\pm\p_\sigma\vec X)^2=(f')^2+f^2$.

Note that $|\vec X|^2,~|\p_\tau\vec X|^2,~|\p_\sigma \vec X|^2=|\vec X|^2$  and $X^0$
are independent of $\sigma$.

\vskip0.3cm

Now let us try to see how the evolution of the above string is modified by  taking into account gravity,
including of course the backreaction.

It appears a very complicated problem, because 
the numerical study of the gravitational collapse is highly nontrivial and took many years 
for realistic cases
(see for instance \cite{Piran,Zwer,Rezzo,Dimme}).  
Here we propose a simplified model based on a consistent truncation of the 
coupled gravitational equations. 

\vskip0.3cm

We begin by considering the standard four dimensional 
Action of  describing gravity and a string in the gravitational field
\footnote{Actually in true String Theory the gravity action should not be "added by hand" to the string dynamics,
because String already contains gravity and the coupled equations should come 
from requiring conformal invariance. However, whereas this has been done 
in a low-energy approximation for
coupling gravity to string fields like the dilaton and $B_{\mu\nu}$,
to our knowledge there is not a similar derivation for the case of a macroscopic string,
which, even worse, would be needed beyond low-energy.
Therefore the above approach of classical equations  is $a~priori$ a heuristic model, 
and thus we feel less unconfortable in making some further truncation.}  
$$
S[g_{\mu\nu}, X^{\mu\nu}]=
-{1\ov\kappa}\int dtd^3x \sqrt{-g}R[g_{\mu\nu}]+
{1\ov\alpha'}\int d\tau d\sigma g_{\mu\nu}[X](\p_\tau X^\mu\p_\tau X^\nu-\p_\sigma X^\mu\p_\sigma X^\nu)
$$
By varying in $g^{\mu\nu}$ and $X^i$, and adding the Virasoro constraint we get:
\be
{1\ov\kappa}\sqrt{-g}R_{\mu\nu}={1\ov\alpha'}\int d\tau d\sigma
\{~ g_{\mu\lambda}g_{\nu\eta}{\cal {T}}^{\lambda\eta}-{1\ov 2}g_{\mu\nu}
~{\cal {T}}^{\lambda\eta}g_{\lambda\eta}~\} ~
\delta(x^0-X^0(\sigma,\tau) )\delta^{3}(\vec x-\vec X(\sigma,\tau))
\label{ein}
\ee
where 
${\cal {T}}^{\lambda\eta}=\p_\tau X^\lambda\p_\tau X^\eta-\p_\sigma X^\lambda\p_\sigma X^\eta$,
$~~$ and
\be
2\p_\tau (g_{ij}\p_\tau  X^j)-2\p_\sigma (\p_\sigma g_{ij} X^j)=~
{\p g_{\lambda\eta}\ov\p x^i}|_{x^\mu= X^\mu}~{\cal {T}}^{\lambda\eta} 
\label{Xeq} 
\ee
plus the constraint equation 
\be
g_{\lambda\eta}(\p_\tau X^\lambda\pm \p_\sigma X^\lambda)(\p_\tau X^\eta\pm \p_\sigma X^\eta)=0
\label{constr}
\ee  

{\bf{2) The truncation.}}
\vskip0.1cm

We truncate the above equations by taking the metric  $g_{\mu\nu}=(g_{00},-1,-1,-1)$
and considering only the $\mu=0,\nu=0$ component of the Einstein equation (\ref{ein}).

In this case we have 
$\sqrt{-g}R_{00}=\sqrt{g_{00}}~{\nabla g_{00}/2}-{(\vec\p g_{00})^2/ 4 \sqrt{g_{00}}}$,
and we get the consistent equations (with $g_s\equiv\kappa/\alpha'$):
\be
\nabla g_{00}=2g_s\int d\tau d\sigma g_{00}^{1/2}~{1\ov 2}\{ g_{00}{\cal{T}}^{00}+
{\cal{T}}^{ij}\delta_{ij}\}~
\delta(x^0-X^0(\sigma,\tau) )\delta^{3}(\vec x-\vec X(\sigma,\tau) )+{(\vec\p g_{00})^2\ov 2g_{00}}
\label{Aeq}
\ee
and
\be
\p^2_\tau \vec X-\p^2_\sigma \vec X=-{1\ov 2}{\p g_{00}\ov\p\vec x}|_{\vec x=\vec X} ((\p_\tau X^0)^2-(\p_\sigma X^0)^2 )
\label{Xt}
\ee
plus the constraint equation 
\be
g_{00}|_{\vec x=\vec X}(\p_\tau X^0\pm \p_\sigma X^0)^2=(\p_\tau\vec X\pm \p_\sigma\vec X)^2
\label{ct}
\ee  

By writing $g_{00}=1-U$, we recognize $U$ to be (twice the negative of) the Newton potential. 
Therefore our truncation corresponds to a Newtonian approximation, in that there are not
retarded effects. Like in many other cases, see for instance \cite{Dimme,Zwer,Dimme2}, and even in cosmology,
the Newtonian approximation  can nevertheless capture
a good part of the physics of the process. 

\vskip0.2cm

In eq.(\ref{Aeq}) the $\tau$-integration gives
$\int d\tau\delta(x^0-X^0)=1/|\p_\tau X^0|$. 

We consistently look for a solution
with the ansatz eq.(\ref{X}) and we will see that on the string $g_{00}[f]\equiv g_{00}[x=X]$ only depends on $f(\tau)$
and $f'(\tau)$. 
\vskip0.2cm
Therefore the constraint eq.(\ref{ct}) gives $|\p_\tau X^0|=R_0 \sqrt{(f')^2+f^2\ov g_{00}[f]}$ and 
$\p_\sigma X^0 =0$. 
\vskip0.2cm
Further, $ {1\ov 2}\{ g_{00}{\cal{T}}^{00}+{\cal{T}}^{ij}\delta_{ij}\}=(R_0 f')^2$.
Thus the equation for $g_{00}$ becomes:
\be
\nabla g_{00}=2 g_s~g_{00}[f]~ R_0{(f')^2\ov\sqrt{(f')^2+f^2}}\int d\sigma
\delta^{3}(\vec x-\vec X(\sigma,\tau) )+{(\vec\p g_{00})^2\ov 2g_{00}}
\label{Aeq2}
\ee

We take the b.c. $\lim_{|\vec x|\to\infty} g_{00}=1$ that is at space infinity
the spacetime metric is flat and the gravitational potential $U$ vanishes. 
Note that for a static string $f'=0$ there is
no gravitational potential $g_{00}=1$, in agreement with a well known effect \cite{Vile}.

\vskip0.2cm

Also, we can rewrite eq.(\ref{Xt}) consistently with our ansatz and the constraint, as
\be
f''+f=-{1\ov 2}{\p g_{00}\ov\p\vec x}|_{\vec x=\vec X} {(f')^2+f^2\ov g_{00}[f]}R_0
\label{Xeq2}
\ee

We are mainly interested on the evolution of the string and to find it we have to compute 
$g_{00}$, and its gradient, on the string, that is on the very location of its source. 
We will neglect the term ${(\vec\p g_{00})^2\ov 2g_{00}}$ on the r.h.s of eq.(\ref{Aeq2}),
since it represents a correction to the source, which is diffused over the space and therefore much less important than the source itself (concentrated on the string).
This is our final truncation.  
 
 \newpage
 
{\bf{3) Solving the equations.} }
\vskip0.2cm
We have now to find $g_{00}$ from eq.(\ref{Aeq2}) (neglecting the last term on the r.h.s.)
and substitute it in eq.(\ref{Xeq2}).

Putting $g_{00}\equiv 1-U$,
we solve eq.(\ref{Aeq2}), avoiding a spurious short-distance log-divergence, which cannot occur 
since String Theory is UV-finite,
 by introducing a minimal cutoff-distance $\sqrt{\alpha'}$ on the string:
$\int d\sigma'/|\vec X(\sigma)-\vec X(\sigma')|\to\int d\sigma'/\sqrt{|\vec X(\sigma)-\vec X(\sigma')|^2+\alpha'}$.
 We get: 
\be
U(\vec x)=2g_s(1-U[f]) {R_0 (f')^2\ov \sqrt{(f')^2+f^2}}\int_0^{2\pi} {d\sigma\ov \sqrt{\vec x^2 +R_0^2f(\tau)^2
-2R_0f(\tau)(xCos(\sigma)+ySin(\sigma))+\alpha'}}
\label{U}
\ee
\vskip0.2cm
We then find  on the string 
$g_{00}[f(\tau)]\equiv g_{00}(\vec x=\vec X(\sigma,\tau))$ (it is $\sigma$ independent):
\be
g_{00}[f]=1-U[f]={1\ov 
1+{2g_s\ov f} ~ {(f')^2\ov \sqrt{(f')^2+f^2}}~H(\alpha'/R_0^2f^2)}
\label{Uf}
\ee
where $H(s)=\int_0^{2\pi} {d\sigma\ov\sqrt{2(1-Cos(\sigma))+s}}$. Note that the only dependence on the 
initial string size $R_0$ is roughly logarithmical, coming from the cutoff.

We take initially $f'(0)=0$ and therefore initially $g_{00}=1$.

We can further evaluate from eq.(\ref{U}) 
$$
{\p g_{00}\ov\p\vec x}|_{\vec x=\vec X}=-{\p U\ov\p\vec x}|_{\vec x=\vec X}=
2g_s~g_{00}[f]~{f\ov R_0 |f|^3}{ (f')^2\ov \sqrt{(f')^2+f^2}}~K(\alpha'/R_0^2f^2)
$$ 
where $K(s)\equiv\int_0^{2\pi} {d\sigma (1-Cos(\sigma))\ov (2(1-Cos(\sigma))+s)^{3/2}}$.

\vskip0.2cm

From that and from eq.(\ref{Xeq2}), we finally find the evolution equation for $f(\tau)$:
\be
f''+f=-g_s~{f\ov |f|^3}(f')^2 \sqrt{(f')^2+f^2}~K(\alpha'/R_0^2f^2)
\label{evol}
\ee
Note that also the evolution depends only logaritmically  on the initial size $R_0$,
through the cutoff funtion $K$. Moreover,  we get just one uncoupled 
evolution equation (\ref{evol}), since $g_{00}[f]$  does not appear explicitly in it.

\vskip0.2cm
We have solved numerically (with Mathematica) the evolution eq.(\ref{evol}), with 
initial conditions $f(0)=1,~f'(0)=0$ taking $g_s=0.01$ and $R_0=10^3 \sqrt{\alpha'}$. 

\vskip0.3cm
{\bf{4) Results describing the collapse.}}
\vskip0.1cm

For a sizable initial interval in $\tau$ the solution looks like the flat one,
but, from a certain moment on, it abruptly accelerates. The numerical programm
stops at $\tau=\tau_0=1.401$ : $~~f'$ and even more dramatically $f''$ show 
a divergent behavior for $\tau\to\tau_0$.
\vskip0.2cm
However the string radius at this final $\tau_0$
is still quite large: $f(\tau_0)=0.053$ and thus $R_0f(\tau_0)=53\cdot\sqrt{\alpha'}$. 

In fig.1 we show $f(\tau)$ (\textcolor{red}{red}) compared with the flat case $f_{flat}$ 
(\textcolor{blue}{blue}) for $0\geq\tau\geq\pi/2$
(the non-flat solution stops at $\tau_0 =1.401$),  $|f'|$ compared with 
$|f'_{flat}|$ and  $f''$ compared with $f''_{flat}$.
\vskip0.5cm

\begin{figure}[ht]\label{fig.1,2,3}
\centerline{
\epsfig{file=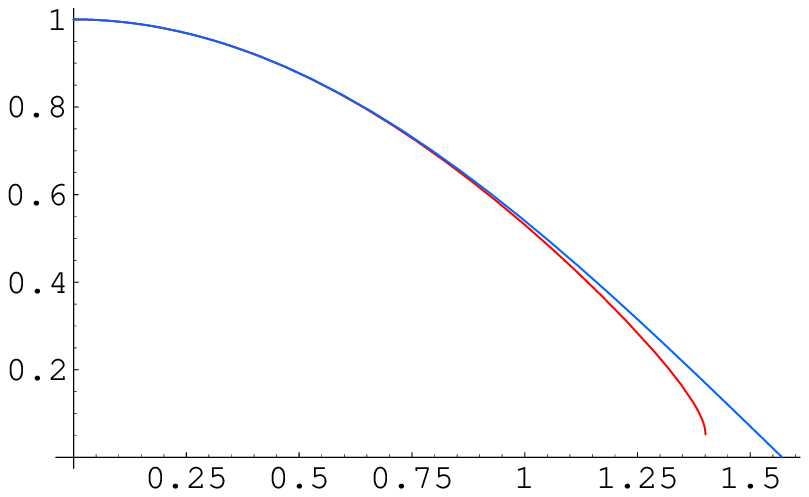,width=.4\textwidth}\
\epsfig{file=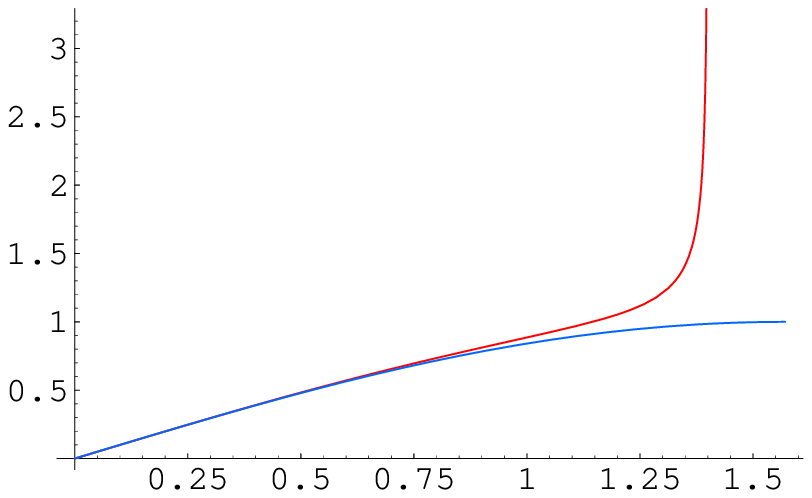,width=.4\textwidth}\
\epsfig{file=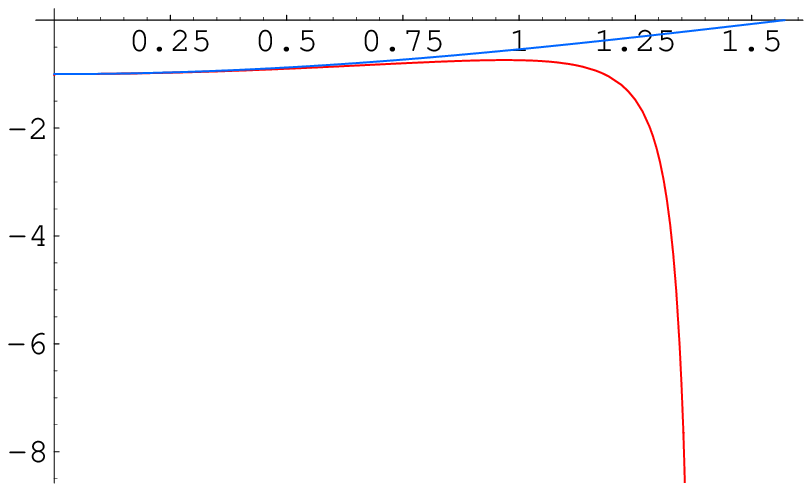,width=.4\textwidth}}
\caption{$f$,$~|f'|$,$~f''$}
\end{figure}

\vskip0.5cm
From eq.(\ref{Uf}) we can also find the behavior of  $g_{00}[f]=1-U[f]$, that is
the $g_{00}$ metric component on the string. 
It stays near 1 untill it rather suddenly decreases, $g_{00}\to 0$
for $\tau\to\tau_0$. We show $g_{00}[f]$ in fig.2. 
\vskip0.2cm

\begin{figure}[ht]\label{fig.4}
\centerline{\epsfig{file=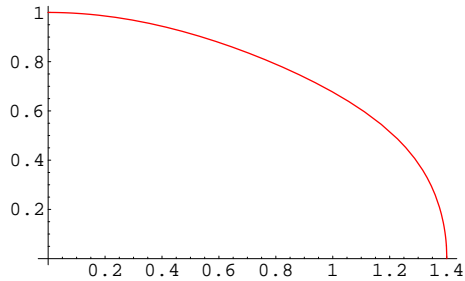,width=.4\textwidth}}
\caption{$g_{00}$ on the string}
\end{figure}

Therefore, since at the endpoint $\tau\to\tau_0$ we get 
$g_{00}\to 0$ (on the string) meaning an infinite redshift for signals coming from the string, 
we can say (amusingly) that 
when the numerical programm stops also the string disappears. 

Note that the infinite red-shift is not due to the string shrinking to zero size (which is not true 
and in any case would not produce any infinity due to the $\alpha'$ cutoff), rather to the 
string velocity approaching the speed of light: $|f'|/\sqrt{(f')^2+f^2}\to 1$, implying the
divergence of the "kinetic energy per unit mass" $\sqrt{(f')^2+f^2}$.

This divergence is due to the non-linearity of the evolution equation (\ref{evol}):
$f$ is almost constant and small near $\tau_0$, and the evolution equation takes
the form $f'' \sim (f')^3$ giving $f'\sim -1/\sqrt{\tau_0-\tau}$.

Note also that when $g_{00}\to 0$ on the string, the curvature component  
$R_{00}\to f(\tau_0) R_0/(2H(\alpha'/R_0^2f(\tau_0)^2)
\int d\sigma\delta^3(\vec x -\vec X(\sigma,\tau))$ 
does not diverge,
apart from being obviously concentrated on the string.

\vskip1cm
{\bf{5) Including the radiation of gravitons.}}
\vskip0.1cm
In the previous discussion we have disregarded the graviton radiation.

Since the evolution is quite violent near the end point $\tau_0$ 
we may suspect that there will be a strong emission of gravitational 
radiation.  This effect can modify quite substantially the picture,
since energy is lost because of radiation. Therefore we can 
expect a dumping due to that dissipation. 

We will now reconsider the evolution by including the effects of the radiation.

\vskip0.2cm
The treatment of the radiation in curved space-time is a very complicated affair (see for instance \cite{Dinv}).
Since we are discussing anyhow a truncated treatment, with the idea of finding 
a simpified physical guideline,  we look for an expression in flat spacetime and
adapt it to our case.     

A convenient formula for the time development of the energy ${\cal{E}}$ of the system,
due to the energy loss by gravitational radiation, is given 
in the Landau-Lifshitz "Classical Field Theory" book \cite{Landau}:

\be
-{d{\cal{E}}\ov dt}|_{rad} =\kappa c_0 \sum_{ij}|{d^3D^{ij}\ov {dt}^{3}}|^2 ~~~~~
 D^{ij}=\int d^3x \rho ~ (3x^ix^j-\vec x^2\delta^{ij})
\label{landau}
\ee 
where $\rho$ is the energy density and $c_0$  a numerical constant. 

In our case, we interpret this formula by taking ${\cal {E}}$ and $\rho$ in the locally inertial frame of the free-fallining string,
and therefore taking their flat-space-time expression, namely:
\be
  {\cal {E}}={1\ov\alpha'}R_0\sqrt{(f')^2+f^2}, ~~ dt=R_0\sqrt{(f')^2+f^2} d\tau ,
   ~~\rho={1\ov\alpha'}R_0\sqrt{(f')^2+f^2}\int d\sigma \delta^3(\vec x-\vec X)
\label{energy}
\ee
giving
\be
D^{ij}={1\ov\alpha'}R_0^3 f^2 \sqrt{(f')^2+f^2} ~ k^{ij}
\ee
where $k^{ij}$ are some constants, for instance $k^{11}=\int d\sigma (3Cos^2 (\sigma) -1)$.
\vskip0.2cm
\vskip0.2cm
In conclusion eq.(\ref{landau}) can be re-written introducing the friction $"Force"~due~ to~ radiation:$ 
\be
f''+f = F_{rad}, ~~~~  F_{rad}=  -g_s c_1 {(f')^2+f^2\ov f'}\cdot |{\cal{D}}^{(3)}|^2
\label{rad1}
\ee
with $c_1$ a numerical constant of order $1$, where we define
\be
{\cal{D}}^{(3)}\equiv {d^3\ov d t^3} (f^2 \sqrt{(f')^2+f^2} )
=({1\ov\sqrt{(f')^2+f^2}}{d\ov d\tau})^3\cdot (f^2 \sqrt{(f')^2+f^2})
\label{D}
\ee
 Now, the previous evolution equation (\ref{evol}) shows  the 
 $"Force"~due~to~selfgravity $:
 \be
f''+f=F_{grav},~~~~~ F_{grav}=-g_s~ {f\ov |f|^3} (f')^2 \sqrt{(f')^2+f^2} ~K(\alpha'/R_0^2f^2)
\label{evolrec}
\ee
Therefore, we obtain the full evolution equation by taking both the 
$"Forces" \\
~due~ to~ selfgravity~AND~radiation$:
\beq
f''+f &=&F_{grav}+F_{rad}, \\ \nonumber
 F_{grav}+F_{rad}&=&-g_s~ {f\ov |f|^3} (f')^2 \sqrt{(f')^2+f^2} ~K(\alpha'/R_0^2f^2)
 -g_s c_1 {(f')^2+f^2\ov f'}~ |{\cal{D}}^{(3)}|^2
\label{pippo}
\eeq 

We see that the radiation contribution has the sign opposite to the one of $f'$, consistently giving an effect of friction.
\vskip0.2cm
This is still a too complicated equation in $f$ for a numerical treatment, because we meet higher derivatives, 
up to $f^{''''}$, appearing in ${\cal{D}}^{(3)}$ .  

In order to proceed, we use the evolution equation
in absence of radiation eq.(\ref{evolrec}) to express $f''$ in terms of $f,f'$, when taking the first derivative
 in the troublesome term ${\cal{D}}^{(3)}$ eq.(\ref{D}), and again when taking the second and the third
\footnote{derivatives of the slowly varying term $K$ are not taken.}.
Thus, in our treatment, the radiation may be somehow under-estimated (because higher derivatives could be much more strong), 
and possibly over-correlated with the self-gravity. 

In this way, the r.h.s. of eq.(19) is expressed in terms of  $f',f$ (we don't show the non-illuminating final form).
We have solved numerically eq.(19),
with the same values of $g_s$ and $R_0$ as in the previous study ignoring radiation (and taking $c_1=1$). 

\vskip0.5cm

{\bf{6) Including radiation: results.}}

\vskip0.2cm

We have first studied the evolution of the string in flat spacetime, but including the radiation, as a kind of benchmark.
In this case, we solve the evolution equation:
\be
f''+f= F_{rad},  ~~~~   F_{rad}=-g_s c_1 {(f')^2+f^2\ov f'}~ |{\cal{D}}^{(3)}|^2 
\label{flat}
\ee
where, similarly to what we said for the non-flat case, we put $f''=-f$ (which holds in absence of radiation), at each step of taking derivatives, in the   troublesome term ${\cal{D}}^{(3)}$.

We show in fig.3  $|f(\tau)|$ from the numerical solution of eq.(\ref{flat}) (with $c_1=1$).
The result makes sense: the dissipation through radiation makes the amplitude of the oscillations of the pulsating string 
smaller and smaller.

\begin{figure}[ht]\label{fig.5}
\centerline{\epsfig{file=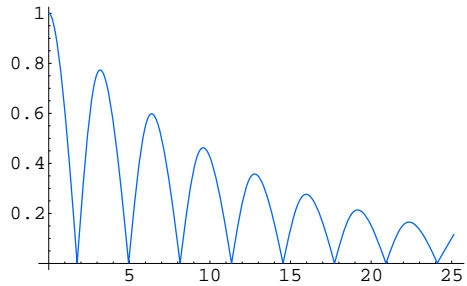,width=.4\textwidth}}
\caption{$|f|$,  flat case}
\end{figure}
\vskip0.2cm
The solution of the case including self-gravity and radiation, eq.(19), shows a  pattern
roughly similar to the flat case.
Now, contrary to when we did not include the radiation, the numerical program does not stop at some $\tau_0$.

At the beginning, the string seems to collapse like when ignoring radiation, but then, contrary to that case, 
the radiation prevents $|f'|$ to diverge to infinity; rather, $|f'|$ passes through a sharp maximum and then goes down 
and up again to another sharp but smaller maximum, and so on.

Remember that the radius of our circular string is $R_0 |f(\tau)|$: now $|f|$  passes through zero near to where there are the maxima of $|f'|$. 
However, $\alpha'$ regulates the self-gravity of the string and therefore there are no singularities 
for $f=0$.  Singularities only occur for $|f'|\to\infty$. 
The consequence of $|f'|$ remaining finite is that $g_{00}$ is never zero, and therefore an infinite gravitational redshift 
on the string does not accur during the evolution.

When $f\to 0$ the result for $g_{00}$ depends on $R_0$ more than logarithmically because $1/|f|\cdot H(\alpha'/(R_0f)^2)\to 2\pi R_0/\sqrt{\alpha'}$  
(see eq.(\ref{Uf})). However, $g_{00}\to 1$ for large $\tau$, with little dependence on $R_0$, because $|f'|\to 0$. 

We have checked that taking $R_0=10^4\sqrt{\alpha'}$, that is an order of magnitude more, gives the same pattern. We have also checked that taking a higher $g_s$, say $g_s=0.1$ or $g_s=1$, makes even larger the
dissipative effect of radiation over attraction.

In fig.4 we show $|f|$ in a first interval (compare with fig.1 where the effect of the radiation is not included), and  also in further larger interval (note the change of scale)
where we see several oscillations with smaller and smaller amplitudes similarly to the flat case.

In fig.5 we show $|f'|$  in a first interval  (compare with fig.1), with a zoom, and in a further larger interval.
Finally, we show  $g_{00}$ in fig.6 (compare with fig.2). 
\begin{figure}[ht]\label{fig.5,6}
\centerline{\epsfig{file=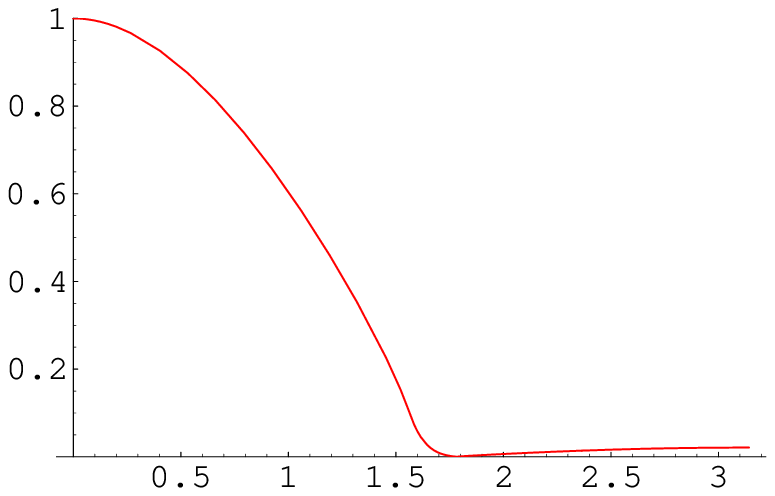,width=.4\textwidth}\
\epsfig{file=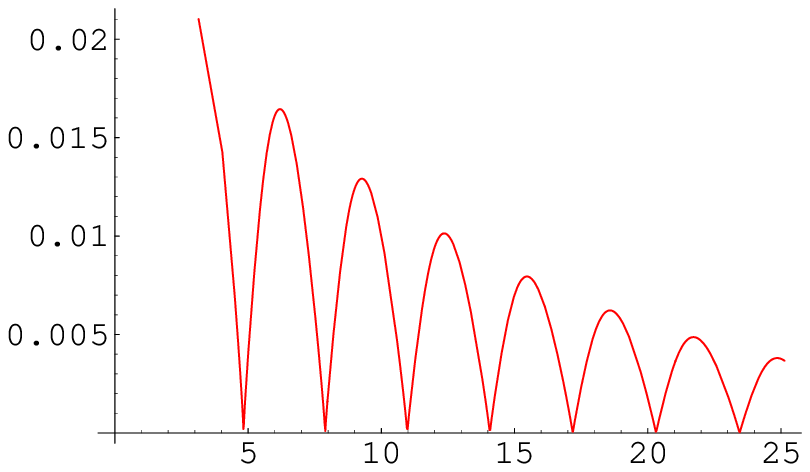,width=.4\textwidth}}
\caption{$|f|$ in two intervals}
\end{figure}
\begin{figure}[ht]\label{fig.7,8,9}
\centerline{
\epsfig{file=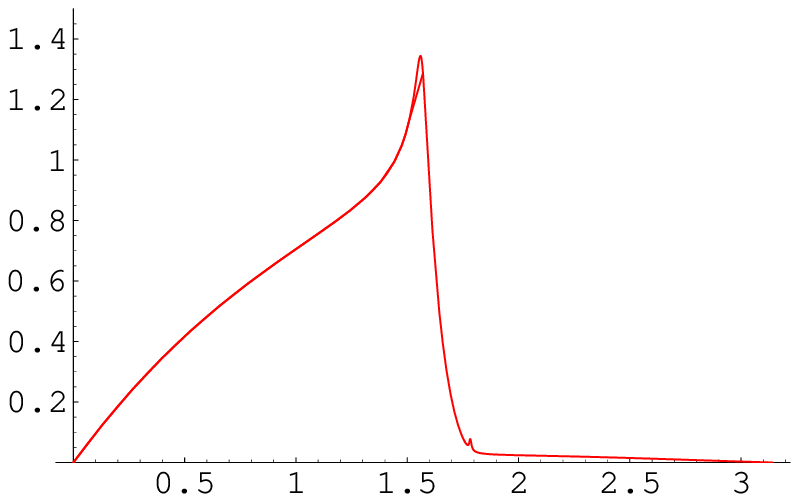,width=.4\textwidth}\
\epsfig{file=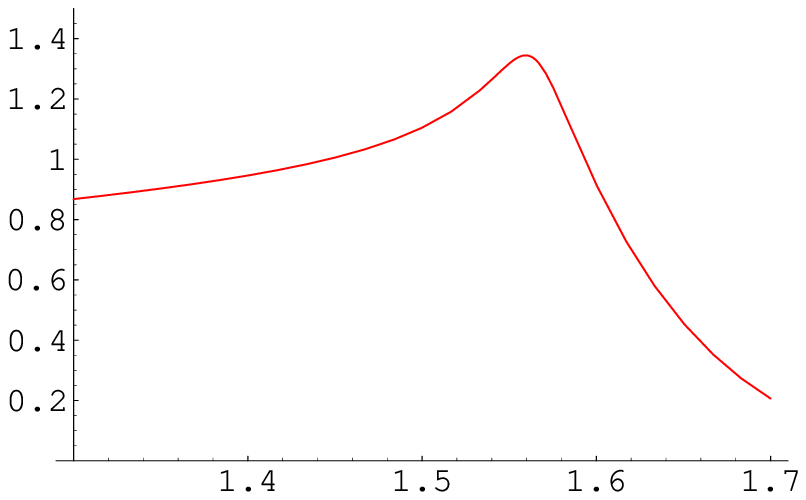,width=.4\textwidth}\
\epsfig{file=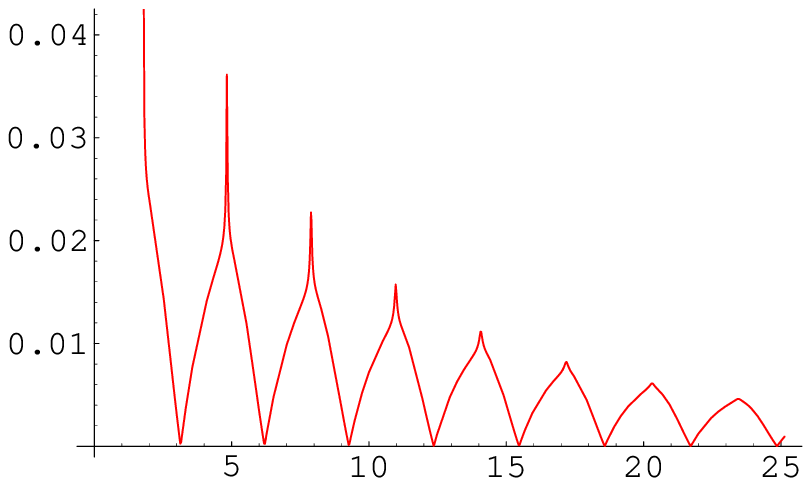,width=.4\textwidth}}
\caption{$|f'|$ in a first interval, with a zoom, and in a further larger interval} 
\end{figure}
\begin{figure}[ht]\label{fig.10}
\centerline{
\epsfig{file=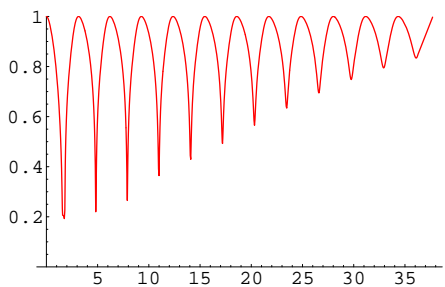,width=.4\textwidth}}
\caption{$g_{00}$ on the string}
\end{figure}

{\bf {7) Conclusions.}}
\vskip0.2cm
The indication is that since radiation prevents the formation of an infinite red-shift,
there is no gravitational collapse making the string disappearing from sight. The string size  periodically
passes through "zero", meaning really $\sim\sqrt{\alpha'}$ ,  without catastrophic  consequences 
(at very small scales the string will pass through a quantum  phase).
The ultimate fate of our string appears to be a quiet shrinking until it completely evaporates by emitting radiation.

In our opinion, this conclusion seems to be reasonable on physical grounds. Of course,
the challenge will be to see, beyond our first attempt, how much our truncated model captures 
the real physics of the string collapse.
\vskip2cm

{\bf{Acknowledgments.}} The author would like to thank J.Russo for many exchanges of ideas in the early stages of this work
and D.Amati for several discussions. Partial support from the EEC network 
MRTN-CT-2004-005104 and the INFN MI12 project is also acknowledged.


\newpage

\begin{thebibliography} {16}

\bibitem{Qua} J.M. Quashnock, D.N.Spergel,
"Gravitational self-interactions of cosmic strings"
Phys.Rev.D {\bf 42}, 2505 (1990).


\bibitem{Vile} A.Vilenkin, E.P.S.Shellard,
"Cosmic strings and other topological defects"
Cambridge University Press (1994).


\bibitem{Hawk2} S.Hawking,
"Gravitational radiation from collapsing cosmic strings"
Phys.Lett. B {\bf 246}, 36 (1990).

\bibitem{Dinv} R. D'Inverno,
"Introducing Einsteins' Relativity"
Clarendon Press (1992).

\bibitem{Landau} L.D.Landau, E.M..Lifshitz,
"The Classical Theory of Fields"
fourth revised english version, Pergamon Press (1975).

\bibitem{Russo} R.Iengo, J.Russo,
"Black Hole formation from collisions of cosmic fundamental strings"
JHEP 0608:079 (2006)
[arXive:hep-th/0606110].

\bibitem{Piran} R.F.Stark, T.Piran,
"Gravitational-wave emission from rotating gravitational collapse"
Phys.Rev.Lett. {\bf 55}, 891 (1985).

\bibitem{Rezzo} L.Baiotti, I.Hawke, L.Rezzolla, E.Schnetter,
"Gravitational-wave emission from rotating gravitational collapse in three dimensions"
Phys.Rev.Lett. {\bf 94}, 131101 (2005)
[arXive:gr-qc/0503016].

\bibitem{Dimme} H.Dimmelmeier, J.A.Font, E.Mueller,
"Relativistic simulation of rotational core collapse. II. Collapse dynamics and gravitational radiation"
Astronomy and Astrophysics {\bf 388}, 917 (2002)
[arXive: astro-ph/0204289].

\bibitem{Zwer} T.Zwerger, E. Mueller,
"Dynamics and gravitational wave signature of axisymmetric rotational core collapse"
Astronomy and Astrophysics {\bf 320}, 209 (1997).

\bibitem{Dimme2} M.Obergaulinger, M.A.Aloy, H.Dimmelmeier, E.Mueller,
"Axisymmetric simulations of magnetorotational core collapse: Approximate inclusion of general relativistic effects"
Astronomy and Astrophysics {\bf 457}, 209 (2006)
[arXive: astro-ph/0602187].

\bibitem{D'Eath} P.D.D'Eath, P.N.Payne,
"Gravitational radiation in black-hole collisions at the speed of light. III. Results and conclusions"
Phys.Rev.D {\bf 46}, 694 (1992).

\bibitem{Giddings} D.M.Eardley, S.B.Giddings,
"Classical black hole production in high-energy collisions"
Phys.Rev.D {\bf 66}, 044011 (2002)
[arXive: gr-qc/0201034].

\bibitem{Giddings2} S.B.Giddings, V.S.Rychkov,
"Black holes from colliding wavepackets"
Phys.Rev.D {\bf 70}, 104026 (2004)
[arXive: hep-th/0409131].

\bibitem{Horo} G.T.Horowitz, J.Polchinski,
"Self-gravitating fundamental strings"
Phys.Rev.D {\bf{57}}, 2557 (1998)
[arXive: hep-th/9707170].

\bibitem{Vene} T.Damour, G.Veneziano,
"Self-gravitating fundamental strings and black holes"
Nucl.Phys.B {\bf{568}}, 93 (2000)
[arXive: hep-th/9907030].

\bibitem{Ama} D.Amati, M.Ciafaloni, G.Veneziano,
"Classical and quantum gravity effects from planckian energy superstring collisions"
Int.Journ.Mod.Phys.A {\bf 3}, 1615 (1988),
"Can spacetime be probed below the string size?"
Phys.Lett.B {\bf 216}, 41 (1989),
"Effective action and all-order gravitational eikonal at planckian energies"
Nucl.PhysB {\bf 347}, 550 (1990).

\bibitem{Ama2} D.Amati,
"The information paradox"
[arXive: hep-th/0612061].


\end{thebibliography}
\end{document}